\documentclass[12pt]{article}
\usepackage{epsfig}
\usepackage{graphicx}
\begin{document}

\def\be{\begin{equation}}
\def\ee{\end{equation}}
\def\bq{\begin{equation}}
\def\eq{\end{equation}}
\def\bqa{\begin{eqnarray}}
\def\eqa{\end{eqnarray}}

\def\roughly#1{\mathrel{\raise.3ex
\hbox{$#1$\kern-.75em\lower1ex\hbox{$\sim$}}}}
\def\lsim{\roughly<}
\def\gsim{\roughly>}
\def\llgm{\left\lgroup\matrix}
\def\rrgm{\right\rgroup}
\def\vectrl #1{\buildrel\leftrightarrow \over #1}
\def\partrl{\vectrl{\partial}}
\def\gslash#1{\slash\hspace*{-0.20cm}#1}

\begin{center}
{\bf Perturbative QCD Evolution and Color Dipole Picture}
\end{center}
\vspace {0.5 cm}
\begin{center}
{\bf Dieter Schildknecht} \\[2.5mm]
Fakult\"{a}t f\"{u}r Physik, Universit\"{a}t Bielefeld \\[1.2mm] 
D-33501 Bielefeld, Germany \\[1.2mm]
and \\[1.2mm]
Max-Planck Institute f\"ur Physik (Werner-Heisenberg-Institut),\\[1.2mm]
F\"ohringer Ring 6, D-80805, M\"unchen, Germany
\end{center}

\vspace{1 cm}

\baselineskip 18pt

\begin{center}
{\bf Abstract}
\end{center}

The proton structure function in the diffraction region of small Bjorken-$x$ and
$10 {\rm GeV}^2 \le Q^2 \le 100 {\rm GeV}^2$ behaves as $F_2 (x, Q^2) = F_2
(W^2) = f_0 \cdot (W^2)^{C_2}$, where $x = Q^2 / W^2$. The exponent $C_2$ of
the $\gamma^* p$ center-of-mass energy squared, $W^2$, is predicted from
evolution of the flavor-singlet quark distribution, $C_2 = 0.29$, and the only
free parameter, the normalization $f_0 = 0.063$, is fitted. The evolution of
the gluon density multiplied by $\alpha_s (Q^2)$ is identical to the evolution
of the flavor-singlet quark density. This simple picture is at variance with the standard
approach to evolution based on the coupled equations of flavor-singlet and
gluon density.

\vspace{1cm}
At sufficiently low values of the Bjorken variable, $x \cong Q^2 / W^2 \le
0.1$, the structure function for deep inelastic electron-proton scattering
(DIS) is in good approximation of perturbative QCD (pQCD) dominated by the
gluon density, or gluon distribution function, of the proton. 

The longitudinal structure function of the proton in this approximation of low
$x$ and reasonably large $Q^2$ is given by \cite{1}
\be
F_L (x, Q^2) = \frac{\alpha_s (Q^2)}{3\pi} \sum^{n_f}_q Q^2_q 6 I_g (x, Q^2)
\label{1}
\ee
with 
\be
I_g (x, Q^2) \equiv \int^1_x \frac{dy}{y} \left( \frac{x}{y} \right)^2 \left( 1
  - \frac{x}{y} \right) G ( y, Q^2), 
\label{2}
\ee
where $G (y, Q^2)\equiv y g (y, Q^2)$,and $g(y, Q^2)$ stands for the gluon density.
The sum over the (active) quark charges squared is denoted by $\sum^{n_f}_q
Q^2_q$. Independently of the specific form of the gluon distribution, for a
wide range of such distributions, the integration in (\ref{2}) yields a
longitudinal structure function directly proportional to the gluon
distribution, but at a rescaled value of $x \rightarrow \xi_L x$, 
\be
F_L (\xi_L x , Q^2) = \frac{\alpha_s (Q^2)}{3\pi} \sum^{n_f}_q Q^2_q G (x ,
Q^2) .
\label{3}
\ee
The rescaling factor in (\ref{3}) has the preferred value of $\xi_L \cong
0.40$.\cite{1}.

The structure function $F_2 (x, Q^2)$ for $x \le 0.1$ in the DIS scheme is
proportional to the flavor-singlet quark distribution, $\sum (x, Q^2)$. For
$n_f = 4$ flavors of quarks, we have 
\be
F_2 (x, Q^2) = \frac{1}{4} \sum^{n_f}_q Q^2_q \cdot x \sum (x, Q^2) =
\frac{5}{18} x \sum (x, Q^2) .
\label{4}
\ee
The evolution of the flavor-singlet quark distribution, and accordingly of $F_2
(x, Q^2)$, with increasing virtuality, $Q^2$, of the photon, $\gamma^*$, is in
good approximation determined by the gluon distribution according to \cite{2,3}
\be
\frac{\partial F_2 (\xi_2 x, Q^2)}{\partial \ln Q^2} = \frac{\alpha_s
  (Q^2)}{3\pi} \sum_q Q^2_q G (x, Q^2).
\label{5}
\ee
In this case of $F_2 (x, Q^2)$ in (\ref{5}), the rescaling factor is given by
$\xi_2 \cong 0.50$. 

By writing
\be
F_L (x, Q^2) = \frac{1}{2\rho + 1} F_2 (x, Q^2), 
\label{6}
\ee  
we introduce the ratio of the structure functions $F_2 (x, Q^2)$ and $F_L (x,
Q^2)$. As long as $\rho$ is allowed to vary with the kinematic variables, $\rho
= \rho (x, Q^2)$, relation (\ref{6}) amounts to a definition of the ratio of
the longitudinal to the transverse photoabsorption cross section,
\be
\frac{\sigma_{\gamma^*_L p} (x,Q^2)}{\sigma_{\gamma^*_T p} (x,Q^2)} = 
\frac{1}{2\rho} . 
\label{7}
\ee
By replacing the right-hand side in (\ref{5}) by (\ref{3}), upon employing the
definition (\ref{6}), we obtain an evolution equation that solely contains the
structure function $F_2 (x, Q^2)$ or, equivalently, the flavor-singlet
distribution (\ref{4}), 
\be
(2\rho + 1) \frac{\partial}{\partial \ln Q^2} F_2 \left( \frac{\xi_L}{\xi_2} x
  , Q^2\right) = F_2 (x, Q^2). 
\label{8}
\ee
Similarly, by replacing the left-hand side in (\ref{6}) by the gluon
distribution according to (\ref{3}), and inserting the result into (\ref{5}),
we find an equation for the gluon distribution that reads 
\be
\frac{\partial}{\partial \ln Q^2} (2\rho + 1) \alpha_s (Q^2) G \left(
  \frac{\xi_2}{\xi_L} x , Q^2 \right) = \alpha_s (Q^2) G (x, Q^2) . 
\label{9}
\ee
Note that, without loss of generality, $\rho = \rho (x, Q^2)$ is allowed, both
in (\ref{8}) and (\ref{9}). 

In the CDP \cite{4}, at sufficiently large $Q^2$, the structure functions
become functions of the $\gamma^* p$ center-of-mass energy, $W$
\cite{5,6}\footnote{The $W$-dependence in (\ref{10}) is a consequence of the
  $W$-dependence of the color-dipole-proton cross section of the CDP
  \cite{5,6}.}
i.e. 
\be F_{2,L} (x, Q^2) = F_{2,L} (W^2 = \frac{Q^2}{x}) . 
\label{10}
\ee
The dependence on the single variable $W^2$ for a wide range of photon
virtualities, $Q^2$, is consistent with the experimental data. Compare fig. 1,
where we show \footnote{We thank Prabhdeep Kaur for providing the plot of the
  experimental data.} the experimental data \cite{7} for $F_2 (x, Q^2)$ as a
function of $1/W^2$ for $10 {\rm GeV}^2 \le Q^2 \le 100 {\rm GeV}^2$. 

In terms of the $W$ dependence from (\ref{10}), the evolution equation
(\ref{8}) reads
\be
(2 \rho_W + 1) \frac{\partial}{\partial \ln W^2} F_2 \left( \frac{\xi_L}{\xi_2}
  W^2 \right) = F_2 (W^2), 
\label{11}
\ee
where the notation $\rho = \rho_W$ indicates that a potential $W$ dependence of
the ratio $\rho$ from (\ref{6}) and (\ref{7}) on the kinematical variables, now
$W^2$, is allowed in (\ref{11}). 

More specifically, we now assume a power law
for the $W$-dependence of $F_2 (W^2)$ in (\ref{10}), 
\be
F_2 (W^2) = f_2 \cdot \left( \frac{W^2}{1 {\rm GeV}^2}\right)^{C_2} = f_2 \cdot
\left( \frac{Q^2}{1 {\rm GeV}^2}\right)^{C_2} x^{-C_2} , 
\label{12}
\ee
where the normalization $f_2$ and the exponent $C_2$ are constants. With
(\ref{12}), the evolution equation (\ref{11}) yields, 
\be
(2\rho_W + 1) C_2 \left( \frac{\xi_L}{\xi_2}\right)^{C_2} = 1 . 
\label{13}
\ee
The conclusion (\ref{13}) from the evolution equation (\ref{11}) clearly rests
on the $W$ dependence of $F_2 (x, Q^2) = F_2 (W^2 = Q^2 / x)$ from the CDP
combined with the power-law ansatz (\ref{12}). According to (\ref{13}), a
constant value of the exponent $C_2$ implies a constant value of $\rho_W = \rho
= {\rm const}$ from (\ref{6}) and (\ref{7}), and vice versa. 

In the CDP, the parameter $\rho$ is associated with the enhanced transverse
size \cite{8,9} of $q \bar q$ fluctuations originating from transversely
polarized photons, $\gamma^*_T$, relative to the transverse size of
$q \bar q$ fluctuations from longitudinally polarized photons,
$\gamma^*_L$. The known distributions of the quark (antiquark) transverse
momentum in transversely versus longitudinally polarized $q \bar q$
fluctuations, via the uncertainty principle, imply an enhanced transverse size
of transversely polarized $q \bar q$ fluctuations. The size enhancement of
definite magnitude yields an enhancement of the transverse relative to the
longitudinal photoabsorption cross section in (\ref{7}) and (\ref{6}) that is
quantitatively fixed by \cite{8,9}
\be
\rho = \frac{4}{3} . 
\label{14}
\ee
We note in passing that the factor 2 in (\ref{7}) is due to the fact that the
intensity of $q \bar q$ pairs from transversely polarized photons in DIS at large $Q^2$ is twice as
large as the one from longitudinally polarized photons. 
This is in distinction from the factor $\rho$ which is a property of the $(q
\bar q)p$ interaction cross section. 

The CDP prediction (\ref{14}), according to (\ref{6}), implies 
\be
F_L (x, Q^2) = 0.27 \cdot F_2 (x, Q^2) . 
\label{15}
\ee
The prediction (\ref{15}) is in agreement with the HERA measurements of $F_L
(x, Q^2)$ for small $x$ and large $Q^2$. Compare refs. \cite{8} and \cite{9}. 

Substituting the empirically verified prediction (\ref{14}) of $\rho = 4/3$
into (\ref{13}) together with the rescaling factors $\xi_L = 0.40$ and $\xi_2 =
0.50$ from (\ref{3}) and (\ref{5}), we find that $C_2$ is determined to be
equal to              
\be
C_2 = \frac{1}{2\rho + 1} \left( \frac{\xi_2}{\xi_L} \right)^{C_2} = 0.29 . 
\label{16}
\ee
We note that the result of $C_2 = 0.29$ is fairly insensitive against variation
of the ratio of the rescaling factors $\xi_2$ and $\xi_L$. The (ad hoc)
variation of this ratio in the interval of $1 \le \xi_2 / \xi_L \le 1.5$ around
the preferred value of $\xi_2 / \xi_L = 1.25$, according to (\ref{16}), yields
$0.27 \le C_2 \le 0.31$\footnote{Note that (\ref{16}) differs from the result in
    ref.\cite{9a} by taking into account the rescaling factor $\xi_L = 0.4$ as
    well as $\rho = 4/3$.}   . Higher energies than the ones that were
  available at HERA are needed for a precision determination of $C_2$ within
  this interval.

Returning to the experimental data for $F_2 = F_2 (W^2 = Q^2 / x)$, in fig. 1,
we show the theoretical result from (\ref{12}) for 
\be
F_2 (W^2) = f_2 \cdot (\frac{W^2}{1{\rm GeV}^2})^{C_2} \equiv 0.063 (\frac{W^2}{1
{\rm GeV}^2})^{0.29} , 
\label{17}
\ee  
where $C_2 = 0.29$ is the theoretical result from (\ref{16}), while the
normalization, $f_2 = 0.063$, was determined by an ``eye-ball'' fit to the experimental data in
fig. 1. 

With only a single fitted parameter, $f_2 = 0.063$, we obtained a
representation of the experimental data for $F_2 (x, Q^2)$ over a wide range of
$10{\rm GeV}^2 \le Q^2 \le 100 {\rm GeV}^2$. A more complete analysis of the
experimental data will be presented in a forthcoming paper \cite{9} where,
within the CDP by refining the previous analysis \cite{5}, 
the extension of the description of the experimental data for
$F_2 (x, Q^2)$ to
$Q^2 = 0$ and $Q^2 > 100 {\rm GeV}^2$ is treated in detail. 

We turn our attention to the evolution of the gluon density in (\ref{9}). For
$\rho = {\rm const}$, compare (\ref{14}), we have from (\ref{9}) 
\be
(2\rho + 1) \frac{\partial}{\partial \ln Q^2} \alpha_S (Q^2) G \left(
  \frac{\xi_2}{\xi_L} x , Q^2 \right) = \alpha_S (Q^2) G (x, Q^2) .
\label{18}
\ee
Comparison of (\ref{18}) with (\ref{8}), taking into account (\ref{4}), reveals
that the evolution of $\alpha_S (Q^2 ) G ( x, Q^2)$, i.e. the evolution of the
gluon density multiplied by $\alpha_S (Q^2)$, coincides with the evolution of
the flavor-singlet quark density \footnote{This result is at variance with the
  usual procedure that supplements the evolution equation for the
  flavor-singlet quark distribution by a separate equation for the gluon
  distribution. See discussion below.}. 

In the language of quark and gluon
distributions, the forward-Compton-scattering amplitude of the CDP in fig.2
corresponds to $\gamma^*$ gluon $\rightarrow q \bar q$ fusion as shown in
fig. 3. According to fig. 3, the evolution of the flavor-singlet quark
distribution induced by the interacting photon, $\gamma^*$, of virtuality $Q^2$
directly measures the evolution of the gluon distribution thus suggesting
identical evolutions of the singlet quark distribution and the gluon
distribution multiplied by the strong coupling $\alpha_S (Q^2)$, as obtained in (\ref{18}). 

Combining (\ref{3}) with (\ref{6}), and taking into account the $W$ dependence
from (\ref{10}), we can directly deduce the gluon distribution from the (fit to
the) experimental data in fig. 1,
\be
\alpha_S (Q^2) G (x, Q^2) = \frac{3\pi}{(2\rho + 1) \sum_q Q^2_q} F_2 \left(
  \frac{1}{\xi_L} W^2 = \frac{Q^2}{\xi_L x} \right),
\label{19}
\ee
where (\ref{17}) is to be inserted on the right-hand side \footnote{Note that
  from (\ref{19}) with (\ref{17}) the gluon distribution function can be
  extracted for any pair of values of $x$ and $Q^2$ with $W^2 = Q^2 / x$ for
  given $W^2$. Even though the underlying relations (\ref{3}) and (\ref{6}) do
  not hold for $Q^2 \le 10 {\rm GeV}^2$, the gluon distribution from (\ref{19})
  remains sensible. The relation of the structure function to the gluon
  distribution becomes modified for $Q^2 \rightarrow 0$, while the
  $W$-dependent gluon distribution remains the one extracted from (\ref{19}),
  compare ref.\cite{9}.We note that our gluon distribution (\ref{19}) is
  manifestly positive in distinction to some result in the literature (compare
  e.g. ref.\cite{10a})}
together with $\rho = 4/3$ from (\ref{14}) and $\sum_q Q^2_q = 10/9$ for four
active flavors.  

Before confronting the gluon distributions obtained from (\ref{19}) with the
results available in the literature \cite{10,11}
we briefly summarize the widely used procedure usually employed when deducing a 
gluon-distribution function from the experimental data on the structure function $F_2
(x, Q^2)$. 

The evolution equation for the flavor-singlet quark distribution is
supplemented by an equation for the gluon distribution, obtained \cite{2} by
replacing the quarks in fig. 3 by (electromagnetically neutral) gluons, disregarding the photon, and
replacing the gluon $\rightarrow$ quark splitting in fig.3 by gluon $\rightarrow$ gluon
splitting. 
The resulting well-known coupled DGLAP equations \cite{2}, in
connection with the fits \cite{11,12,13,14} to the experimental data on $F_2 (x,
Q^2)$, are then solved numerically. 

The DGLAP gluon-evolution equation has an approximate analytic solution, the well-know
double-asymptotic solution (DAS) \cite{15} that is understood \cite{2,4} as
resummation of a gluon ladder subject to certain ``ordering assumptions'' on
the gluon momenta, compare fig.4. The DAS corresponds \cite{4} to introducing
an $x$-dependent generalized dipole cross section into the CDP, thus resolving
the lower blob in fig.2. The $x$-dependence of the DAS of the DGLAP approach is
at variance with our requirement of a $W$-dependent color-dipole cross
section. It is precisely this $W$ dependence that (within the CDP) allows
\cite{5,9} for a transition to the region of $Q^2 \rightarrow 0$, including
photoproduction by real photons. 

Based upon an analysis employing the DAS of the DGLAP equations, the exponent,
$\lambda$, of the Bjorken-$x$ dependence of the gluon distribution, $G (x, Q^2) \sim
x^{-\lambda}$, was predicted \cite{16} as $\lambda = 0.32 \pm 0.05$. This
result is consistent with our prediction of $C_2 = 0.29$ from (\ref{16}),
(\ref{17}) and (\ref{19}). 

Application of DGLAP evolution to the ``hard Pomeron'' part of a Regge fit
\cite{11} to the experimental data for $F_2 (x, Q^2)$, led to $G(x,Q^2) \sim
x^{-\epsilon_0}$, where $\epsilon_0 = 0.427$ is the fit parameter
characterizing the (necessary) hard Pomeron contribution to the structure
function, $F_2 (x, Q^2) \sim x^{-\epsilon_0}$. While this $x$ dependence is
somewhat stronger, the $Q^2$ dependence of the gluon distribution extracted
\cite{11} from the Regge fit is somewhat weaker than ours that coincides with
the $x$ dependence and is determined by $\alpha_S (Q^2) G(x,Q^2) \sim (Q^2)^{C_2} =
(Q^2)^{0.29}$. 

The most elaborate and technically demanding numerical extractions of
valence-quark as well as sea-quark and gluon distributions were carried out by
so-called global fits to the experimental data of the structure function $F_2
(x, Q^2)$ by several collaborations \cite{12,13,14}. 
Comparing the results of the different collaborations collected in the Durham
Data Base \cite{10}, one finds a significant spread of the values of the
extracted gluon distributions as a function of $x$ as well as  $Q^2$. 

Our results for the gluon distribution from (\ref{19}) with (\ref{17}) lie
within the range of the distributions from the hard Pomeron \cite{11} and from
refs.\cite{12,13,14} as given by the Durham Data Base \cite{10}. More details
will be presented in \cite{9}. 

In summary:
\begin{itemize}
\item[i)]
By combining pQCD in the approximation (\ref{3}) and (\ref{5}) with the
$W$-dependence of $F_2 (x, Q^2) = F_2 (W^2 = Q^2 / x)$ from the CDP, we find
that the evolution equation for the flavor-singlet quark distribution predicts
the exponent $C_2 = 0.29$, where $F_2 (W^2) \sim (W^2)^{C_2}$. Only one fitted
parameter, the normalization of $F_2 (W^2)$, is required to represent $F_2
(W^2)$ for $x < 0.1$ in the wide range of $10 {\rm GeV}^2 \le Q^2 \le 100 {\rm
  GeV}^2$. Our results for $F_2 (x, Q^2) = F_2 (W^2)$ within the CDP of pQCD can be
smoothly extended to $Q^2 \rightarrow 0$. 

\item[ii)]
The evolution of the gluon-distribution function multiplied by $\alpha_S
(Q^2)$ is identical to the flavor-singlet evolution. This result is at
variance with the results from the usual extraction of the gluon distribution
that relies on supplementing the DGLAP evolution of the flavor-singlet quark
distribution by the gluon-evolution equation. 

\item[iii)]
Our gluon distribution is directly related to (the fit to) the experimental
data for $F_2 (W^2)$ by a known proportionality constant. Our extraction of the (manifestly
positive) gluon
distribution from the experimental data is transparent, as far as the
underlying theoretical assumptions and the relation to the experimental data 
are concerned, and it is straight forward and simple. 

\item[iv)]
The results for the gluon distribution from the analysis of the experimental
data by different collaborations differ significantly from each other. 
Within this wide range, our gluon
distribution is compatible with the published ones, even though both, our
underlying assumptions and our procedure, differ appreciably from the ones in
the literature.
\end{itemize} 

\vspace{1cm}\noindent
{\bf Acknowledgement}

\medskip\noindent
I am grateful to Prabhdeep Kaur for providing the plot of the experimental data
for the structure function.

\vspace{1cm}

\begin{figure}[htb]
\centerline{\epsfig{file=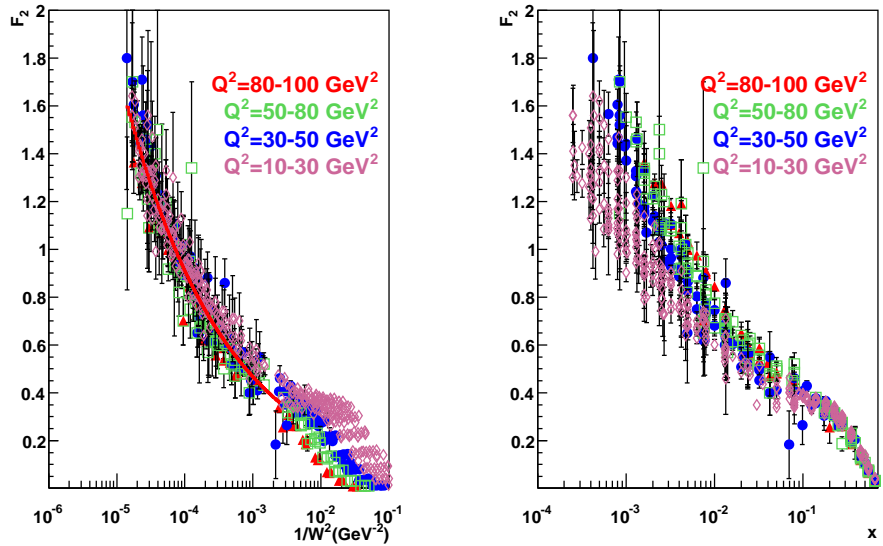,width=13cm}}
\caption{The experimental data on the proton structure function $F_2 (x, Q^2)$ as a
function of $1 / W^2$. The theoretical curve is based on (\ref{17}). For
comparison, we also show $F_2 (x, Q^2)$ as a function of $x \cong Q^2 / W^2$.}
\end{figure}

\vspace*{2cm}

\begin{figure}[htb]
\centerline{\epsfig{file=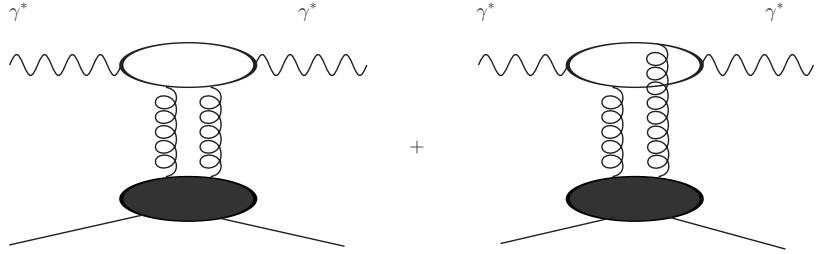,width=11cm}}
\caption{The forward Compton amplitude of the CDP}
\end{figure}

\vspace*{2truecm}

\begin{figure}[htb]
\centerline{\epsfig{file=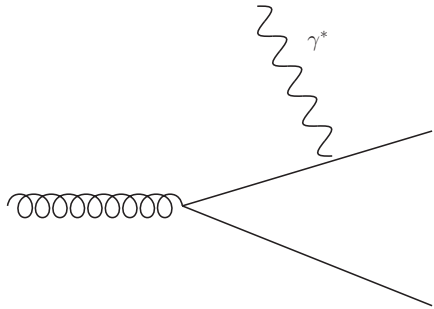,width=6cm}}
\caption{Photon-gluon $\rightarrow q \bar q$ fusion, equivalent to the CDP from
fig.1.}
\end{figure}

\vspace*{2truecm}

\begin{figure}[htb]
\centerline{\epsfig{file=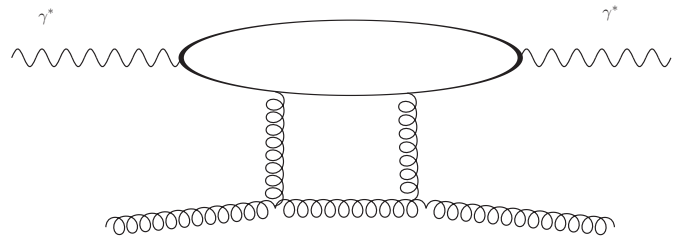,width=9cm}}
\caption{Higher order corrections to photon-gluon $\rightarrow q \bar q$ fusion
resolving the lower blob in fig.2. The lower part of the diagram must be
extended to become a gluon ladder.}  
\end{figure}



\begin{thebibliography}{99}
\bibitem{1}
A.D. Martin et al., Phys. Rev. D37 (1988) 1161; \\
A.M. Cooper-Sarkar et al., Z. Phys. C39 (1988) 281. 

\bibitem{2}
L.N. Lipatov, Sov. J. Nucl. Phys. 20 (1975) 95; \\
V.N. Gribov and L.N. Lipatov, Sov. J. Nucl. Phys. 15 (1972) 438; \\
G. Altarelli and G. Parisi, Nucl. Phys. B126 (1977) 298; \\
Yu. L. Dokshitzer, Sov. Phys. JETP 46 (1977) 641.

\bibitem{3}
K. Prytz, Phys. Lett. B311 (1993) 286.

\bibitem{4}
N.N. Nikolaev, B.G. Zahkarov, Z. Phys. C 49 (1991) 607; 
Z. Phys. C64 (1994) 631.

\bibitem{5}
D. Schildknecht, Contribution to Diffraction 2000, Centraro, Italy, September
2-7, 2000, Nucl. Phys. B  (Proc. Supplement) 99 (2001) 121; \\
D. Schildknecht, B. Surrow, M. Tentyukov, Phys. Lett. B499 (2001) 116; \\
G. Cvetic, D. Schildknecht, B. Surrow, M. Tentyukov, EPJC 20 (2001) 77; \\
D. Schildknecht, B. Surrow, M. Tentyukov, Mod. Phys. Lett. A16 (2001) 1829.

\bibitem{6}
C. Ewerz and O. Nachtmann, Annals of Physics 322 (2007) 1635; 322 (2007) 1670;
\\
C. Ewerz, A. v. Manteuffel, O. Nachtmann, arXiv:1101.0288 [hep-ph].

\bibitem{7}
Durham Data Base, http://durpdg.dur.ac.uk/HEPDATA/REAC

\bibitem{8}
M. Kuroda and D. Schildknecht, Phys. Lett. B670 (2008) 129. \\
D. Schildknecht, Contribution to DIS2009, Madrid, April 25 to 30, 2009,
arXiv0907.0545 [hep-ph]

\bibitem{9}
M. Kuroda and D. Schildknecht, in preparation

\bibitem{9a}
M. Kuroda and D. Schildknecht, Phys. Lett. B618 (2005) 84.

\bibitem{10a}
A. Cooper-Sarkar, arXiv:0901.4001 [hep-ph].

\bibitem{10}
Durham Data Base, http://durpdg.dur.ac.uk/HEPDATA/PDF.

\bibitem{11}
A. Donnachie and P.V. Landshoff, Phys. Lett. B533 (2002) 277,hep-ph/0111427;  
Acta Physica Polonica B34 (2003) 2989, hep-ph/0305171.

\bibitem{12}
M. Gl\"uck, E. Reya, A. Vogt, Z. Phys. C67 (1994) 433; Eur. Phys. J. C5 (1998) 461.

\bibitem{13}
A.D. Martin, R.G. Roberts, W.J. Stirling and R.S. Thorne, Eur. Phys. J. C18
(2000) 117.

\bibitem{14}
CTEQ collaboration: J. Pumplin et al., JHEP 0207 (2002) 012.

\bibitem{15}
A. De Rujula et al., Phys. Rev. D10 (1974) 1649;\\
R.D. Ball and S. Forte, Phys. Lett. B335 (1994) 77.

\bibitem{16}
F. Caola and S. Forte, arXiv:0802.187802 [hep-ph], Phys. Rev. Letters 101
(2008) 022001.



\end{thebibliography}
\end{document}